# Dynamic Origin of Vortex Core Switching in Soft Magnetic Nanodots

Konstantin Yu. Guslienko, Ki-Suk Lee, and Sang-Koog Kim*

*Research Center for Spin Dynamics & Spin-Wave Devices, Seoul National University, Seoul 151-744, Republic of Korea*
*Nanospintronics Laboratory, Department of Materials Science and Engineering, College of Engineering, Seoul National University, Seoul 151-744, Republic of Korea*

## Abstract

The magnetic vortex with the in-plane curling magnetization and the out-of-plane magnetization at the core is a unique ground state in nanoscale magnetic elements. This kind of magnetic vortex can be used as a memory unit for information storage, through its downward or upward core-orientation and, thus, controllable core switching deserves some special attention. Our analytical and micromagnetic calculations reveal that the origin of the vortex core reversal is a gyrotropic field. This field is induced by vortex dynamic motion and is proportional to the velocity of the moving vortex. Our calculations elucidate the physical origin of the vortex core dynamic reversal and offer a key to effective manipulation of the vortex-core orientation.

Vortex patterns exist in physical dynamic systems of greatly varying temporal and spatial scales, ranging from quantized vortices in superfluids and superconductors [1] to water whirlpools, atmospheric tornadoes, and galaxies of the universe. Magnetic vortices are among the most prominent examples of the magnetization (**M**) ground states typically observed in magnetic particles of submicron size, such as nanodots [2-4]. The control of the **M** reversal in small particles is on the cutting edge of modern nanomagnetism. Vortices are elementary objects that describe the **M** reversal in such particles via their nucleation, propagation and annihilation [5]. The magnetic vortex consists of the core (VC), a small area of 10 - 20 nm radius $R_c$, where **M** deviates from the dot plane [2,3], and the main part with the in-plane curling **M** around its core area. Existence of the VC non-zero topological charges [1,6] in combination with the long-range magnetostatic interaction leads to unique effects in vortex dynamics. Non-trivial vortex excitations emerge that exist neither in bulk magnetic systems nor in continuous films. The magnetic vortex in nanodots possesses, in particular, a dynamic excitation that corresponds to the rotation of its core around an equilibrium position at a characteristic frequency of several hundred MHz [7-12]. The unique nature of the vortex **M** distribution is expected to enable applications in magnetic data storage [13-16], spin wave generation [17], and others. For instance, a magnetic vortex can be used as an information carrier because it has two discrete states of core orientation (up and down) and two directions of in-plane **M** rotation.

Very recently, the VC switching in magnetic nanodots (see Fig. 1) has been experimentally and numerically demonstrated using a small-amplitude field pulse [13,18-20] as well as a spin-polarized current [14]. It was supposed that the switching occurs through vortex-antivortex pair creation and annihilation [13,18-20]. However, despite the considerable interest in dynamic VC switching, the origin of this new phenomenon has not been unveiled. However, in this Letter, we are able to explain the physical origin of the VC orientation reversal. Before we knew only how the reversal takes place, but now we can also describe why it occurs.

In order to explain the physical origin of the VC orientation reversal, we consider the dynamic vortex **M** distribution $\mathbf{m}(\mathbf{r},t) = \mathbf{M}(\mathbf{r},t)/M_s$, where $M_s = |\mathbf{M}(\mathbf{r},t)|$ is the saturation magnetization, and **r** is the coordinate in the dot plane. We represent the time-derivative term in the magnetization equation of motion (the Landau-Lifshitz equation; see Ref. 21) by a "gyrotropic" field **h**. This is a very useful way to describe the VC switching. The gyrotropic field **h** and corresponding torque can be calculated explicitly for a moving magnetic vortex. The **h** is proportional to the vortex velocity **v** or to the shift of the VC position **X**(*t*) from the equilibrium position in nanodot. The equation of the magnetization **M** motion can be formulated, in general, via an effective field, which is variational derivative of the corresponding Lagrangian $L = \int d^3\mathbf{x}\ell(\mathbf{m},\dot{\mathbf{m}})$ with respect to **m** [21]. The Lagrangian density $\ell$ is equal to $g - w$, where $g$ is the kinetic (gyrotropic) term containing the time derivative of magnetization $\dot{\mathbf{m}}$, and $w$ is the magnetic energy density. To find the kinetic part of the effective field, which we call *gyrotropic field* or *gyrofield*, we use the kinetic part of the Lagrangian density [21] in the form:

$$g = \frac{M_s}{\gamma} \frac{\mathbf{n} \cdot (\mathbf{m} \times \dot{\mathbf{m}})}{1 + \mathbf{m} \cdot \mathbf{n}}, \qquad (1)$$

where $\gamma$ is the gyromagnetic ratio, and the unit vector $\mathbf{n}$ describes the VC magnetization orientation. The Lagrangian, Eq. (1), has some singularity along the line $\mathbf{m} \cdot \mathbf{n} = -1$, which corresponds to Dirac's string along the direction $\mathbf{n}$ being related to the magnetic monopole [22], which breaks rotation invariance. To make this point clearer we temporarily use the variable length magnetization $\mathbf{M}' = M\mathbf{m}$ (without constraint $|\mathbf{M}'| = M_s$). Then, we can write $g = \mathbf{A} \cdot \dot{\mathbf{M}}'$, where the variable $\mathbf{M}'$ has sense of a charged particle coordinate and the vector potential of effective magnetic field is $\mathbf{A} = (M_s/\gamma)[\mathbf{n} \times \mathbf{M}']/M(M + \mathbf{n} \cdot \mathbf{M}')$ [23]. This potential is singular along the line $\mathbf{M}' \cdot \mathbf{n} = -M$. The effective magnetic field $\mathbf{B} = \nabla_{\mathbf{M}'} \times \mathbf{A} = (M_s/\gamma)\mathbf{M}'/M^3$ describes a magnetic monopole in $\mathbf{M}'$-space with the "magnetic" charge $M_s/\gamma$ located at the origin $\mathbf{M}' = 0$. Then we interpret a magnetic vortex as crossing of the Dirac's string with surface of the sphere $|\mathbf{M}'| = M_s$ in $\mathbf{M}'$-space. This leads to the kinetic part of the Lagrangian in the form of Eq. (1).

We choose $\mathbf{n} = \hat{\mathbf{z}}$ for the magnetic vortex with the core magnetization directed perpendicularly to the nanodot plane (Fig. 1). It is easy to check that in the angular representation of $\mathbf{m}$ via spherical angles $\Theta$, $\Phi$, Eq. (1) reduces to the well known form $g = (M_s/\gamma)(1 - \cos\Theta)\dot{\Phi}$ [21]. Let define the VC polarization $p$ as sign of the vortex $m_z$ component in its extreme point (the core magnetization orientation) [1,6]. We use Eq. (1) in the form $g = (M_s/\gamma)(\mathbf{m} \times \dot{\mathbf{m}})_z /(p + m_z)$, which corresponds to the Dirac's string located along the negative/positive $Oz$ semi-axis for $p$=+1/-1. The gyrofield is then defined as $\mathbf{h} = M_s^{-1} \partial g / \partial \mathbf{m}$, and its z-component interacting with the VC can be expressed as

$$h_z = -\frac{1}{\gamma} \frac{(\mathbf{m} \times \dot{\mathbf{m}})_z}{(m_z + p)^2}. \qquad (2)$$

Eq. (2) in the angular representation is reduced to $h_z = -\tan^{2p}(\Theta/2)\dot{\Phi}/\gamma$ and the string is located along the line $\Theta = \pi$ or $\Theta = 0$ for $p = \pm 1$, respectively. The gyrofield $\mathbf{h}$ is time dependent and, in particular, for a moving vortex $\mathbf{m}(\mathbf{r},t) = \mathbf{m}(\mathbf{r} - \mathbf{X}(t))$, is proportional to the vortex velocity $\mathbf{v} = \dot{\mathbf{X}}$ introduced by the equation $\dot{\mathbf{m}} = -(\mathbf{v} \cdot \nabla)\mathbf{m}$. Its z-component (perpendicular to the dot plane) is the origin of the vortex profile $m_z(\mathbf{r},t)$ dynamical deformation [12] (See Fig. 2 and Supplementary Movies 1 and 2 [24]). That deformation, leading eventually to the VC reversal and the VC velocity drop (shown in Fig. 3), was calculated numerically in Ref. 14. The calculated **h** and torque [17] are concentrated near the VC, where the vortex magnetization component $m_z$ essentially differs from zero. That is, the dynamic core profile deformation up to the VC reversal is caused by the "effective" Zeeman energy, $w_z(t) = -h_z(t) m_z(t)$, which is indeed the driving force of the process. The **h** deforms the vortex magnetization distribution $m_z(\mathbf{r},t)$, creating a dip on the inner side of the core to match the field profile $h_z(\mathbf{r},t)$ (Fig. 2). Assuming a circular VC motion of frequency $\boldsymbol{\omega} = \omega \hat{\mathbf{z}}$ driven by an external in-plane oscillating field (Figs. 1, 3) in a circular nanodot, the equation $\mathbf{v} = \boldsymbol{\omega} \times \mathbf{X}$ holds [25] and the z-component of the gyrofield is $h_z(\mathbf{r},t) = (\omega/\gamma) F(r)(\hat{\mathbf{r}} \cdot \mathbf{X}(t))/R_c$ (here $\hat{\mathbf{r}}$ indicates the unit vector along **r**) within the linear approximation on **X**. The spatial distribution of the field reveals different $h_z$ signs depending on the core coordinate. The gyrotropic field changes sign on the line perpendicular to **X**, as shown in Fig. 2. The radial function $F(r)$ depends on the detailed VC profile and reaches its maximum $F(r) \approx 1$ within the VC, $r < R_c$. This allows for the estimation of the maximum value of the gyrofield as $h_z \approx (\omega/\gamma)|\mathbf{X}|/R_c$. This value is about 1 kOe near the VC, and is sufficient to create an essential deformation of the VC magnetization profile, whereas $h_z$ outside the core is ~ 10 Oe only. Such a large and spatially non-uniform gyrofield interacting with the VC leads to its significant deformation and, eventually, to the core orientation switching at some critical value $h_c \sim v_c$ (where $v_c$ is the critical vortex velocity, see Fig. 3). The vortex velocity $\upsilon$ is proportional to the AC field or spin-polarized current amplitude, whereas the critical velocity $\upsilon_c$ is an intrinsic parameter for the given nanodot geometry and does not depend on the driving force parameters. We estimate $\upsilon_c$ as $\upsilon_c \approx \gamma M_s R_c \approx \gamma(2A)^{1/2}$ ($A$ is the exchange stiffness), that is about 320 m/s for the Permalloy nanodot. This estimation of $\upsilon_c$ is in good agreement with numerical simulations [26] as shown in Fig. 3(b). In some sense, the $\upsilon_c$ is similar to the Walker critical velocity for one dimensional domain walls [21] because it corresponds to the transition from the linear gyrotropic motion to the strongly nonlinear regime of the VC orientation oscillations. Under the influence of an AC driving field, the VC amplitude $|\mathbf{X}(t)|$ and velocity increase with time (see Fig. 3(b), (c) and Supplementary Movie 3 [24]) and when they reach some critical values the VC suddenly reverses its polarization. That is, the moving vortex adsorbs the energy of the oscillating driving field and increases its profile deformation and velocity. Then, the energy excess is emitted in the form of spin waves [20] and the vortex velocity and oscillation amplitude sharply decrease [Fig. 3(b) and (c)]. In the case of periodic driving field this VC reversal process also repeats periodically. This effect was numerically observed in Ref. 20 and was used to generate the strong spin waves immediately following the VC reversal.

To confirm the physical mechanism of the VC reversal as driven by an oscillating magnetic field, we conducted micromagnetic calculations [26,27] and showed that the spatial distribution of $h_z$ (Fig. 2) is indeed close to the distribution of $m_z(r,t)$. For comparison, the vortex dynamic **M** distribution is plotted in Fig. 2. This figure clearly illustrates that the VC dynamic distortion profile follows the distribution of the z-component of the dynamic gyrotropic field. If we consider the time evolution of the extreme value of the gyrofield [Fig. 3(d)] and the out-of-plane **M** component $m_z$ [Fig. 3(e)], it is also evident that the component $m_z(t)$ practically follows the dependence $h_z(t)$. The gyrofield $h_z$ reaches infinite values in the time moments of the VC reversal [see Fig. 3(d)] due to our definition of this field by Eq. (2). The consideration of the gyrofield above is valid for any static VC profile as well as for any ansatz of the **M** distribution of the moving vortex in the form $\mathbf{m}(\mathbf{r},t) = \mathbf{m}(\mathbf{r}, \mathbf{X}(t))$. The distortion of the moving VC profile due to the gyrofield, and the resulting VC orientation reversal are general phenomena for any dot shape. That dynamical distortion was observed, in particular, in square dots [13] and circular dots [14] excited from the centered magnetic vortex ground state.

The results here presented offer insight into the vortex core reversal as driven by either a small amplitude in-plane oscillating magnetic field or a spin-polarized current. The understanding of the physical origin of vortex core switching is the key to the effective manipulation of the dynamic switching of core orientation via the choice of appropriate strengths and frequencies of driving forces as well as of magnetic nanoelement geometry.

**Acknowledgements**

This work was supported by Creative Research Initiatives (Research Center for Spin Dynamics & Spin-Wave Devices) of MOST/KOSEF.

# References

*Corresponding author, electronic address: sangkoog@snu.ac.kr

[1] N. Manton and P. Sutcliffe, *Topological solitons* (Cambridge Univ. Press, Cambridge, 2004), Chapt. 1.

[2] T. Shinjo *et al.*, Science **289**, 930 (2000).

[3] A. Wachowiak *et al.*, Science **298**, 577 (2002).

[4] R. P. Cowburn *et al.*, Phys. Rev. Lett. **83**, 1042 (1999).

[5] K. Yu. Guslienko *et al.*, Phys. Rev. B **65**, 024414 (2002).

[6] A. M. Kosevich, B. A. Ivanov, and A.S. Kovalev, *Phys. Rep.* **194**, 117 (1990).

[7] K. Yu. Guslienko *et al.*, J. Appl. Phys. **91**, 8037 (2002).

[8] J. P. Park *et al.*, Phys. Rev. B **67**, 020403(R) (2003).

[9] S. B. Choe *et al.*, Science **304**, 420 (2004).

[10] X. Zhu *et al.*, *Phys. Rev. B* **71**, 180408 (2005).

[11] K. Yu. Guslienko *et al.*, Phys. Rev. Lett. **96**, 067205 (2006).

[12] V. Novosad *et al.*, Phys. Rev. B **72**, 024455 (2005).

[13] B. Van Waeyenberge *et al.*, Nature **444**, 461 (2006).

[14] K. Yamada *et al.*, *Nature Mater.* **6**, 269 (2007).

[15] R. P. Cowburn, *Nature Materials* **6**, 255 (2007).

[16] J. Thomas, *Nature Nanotechnology* **2**, 206 (2007).

[17] S. Choi, K.-S. Lee, K. Yu. Guslienko, and S.-K. Kim, Phys. Rev. Lett. **98**, 087205 (2007).

[18] R. Hertel, S. Gliga, M. Fahnle, and C.M. Schneider, Phys. Rev. Lett. **98**, 117201 (2007).

[19] Q.F. Xiao et al., Appl. Phys. Lett. **89**, 262507 (2006).

[20] K.-S. Lee, K. Y. Guslienko, J.-Y. Lee, and S.-K. Kim, arXiv:cond-mat/0703538v1.

[21] A. Hubert and R. Schafer, *Magnetic Domains*, Sec. 3.6.6 (Springer-Verlag, Berlin, 1998).

[22] F.D.M. Haldane, *Phys. Rev Lett*. **57**, 1488-1491 (1986).

[23] E. Witten, *Nucl. Phys*. **B223**, 422 (1983); B. A. Ivanov and N. E. Kulagin, *J. Exp. Theor. Phys*. **99**, 1291 (2004).

[24] See EPAPS Document No. XXX for the supplementary three movie files. For more information on EPAPS, see http://www.aip.org/pubservs/epaps.html.

[25] The typical VC dynamic shift in submicron size dots is $|\mathbf{X}|$ ~100 nm, and the vortex velocity is about 100 m/s, which is comparable to the velocity of moving domain walls.

[26] We used the following material and geometrical parameters for the Permalloy nanodot: $M_s$ = 8.6 x $10^5$ A/m, the exchange stiffness $A$ = 1.3 x $10^{-11}$ J/m, the damping constant $\alpha$ = 0.01, $\gamma$ = 2.21 x $10^5$ m/As, the dot radius of 150 nm, and thickness of 20 nm (Fig. 1). The vortex motion was driven by an external oscillating in-plane field with linear polarization $\mathbf{H}(t) = H_0 \sin(\omega t)\hat{\mathbf{y}}$ with the amplitude $H_0$ = 50 Oe. In order to effectively excite the vortex gyrotropic motion we

chose a particular value of $\omega/2\pi$ = 580 MHz, which is equal to the vortex eigenfrequency for the given dot material and geometry. The critical vortex velocity was numerically calculated to be $v_c$ = 350±20 m/s for the particular magnetic nanodot parameters.

[27] M. Donahue and D. Porter OOMMF User's Guide, version 1.0. Interagency Report NISTIR 6376 (National Institute of Standards and Technology, Gaithersburg, Maryland, 1999).

**Figure captions**

**FIG. 1** (color online). Geometry of the model and scheme of the VC reversal in a cylindrical magnetic nanodot. The in-plane **M** distribution is marked by different colors, as indicated by the color wheel. The spikes correspond to the core magnetization: the switching occurs from the "upward" (left) to "downward" orientation (right). The oscillating in-plane magnetic field amplitude is $H_0$ = 50 Oe, and the field frequency is $\omega/2\pi$ = 580 MHz.

**FIG. 2** (color online). Snapshot images at the selected moments of the spatial **M** distribution and of the gyrofield distribution for a moving vortex. The color of the surfaces indicates the local out-of-plane normalized magnetization $m_z$ in the first row and the local out-of-plane component $h_z$ in the second row. The streamlines with the small arrows indicate the in-plane direction of the local **m**. The VC magnetization (red) and the gyrofield z-component (green) profiles along the line crossing the VC and the nanodot center are shown in the third row. The static vortex profile (black dashed line) was intentionally shifted for direct comparison with the dynamic vortex profile (red solid line).

**FIG. 3** (color online). (a) The simulated trajectories of the VC motion under an oscillating magnetic field during the time periods of $t = 0 - 8$ ns for the upper panel (shaded circle) and $t = 8 - 16$ ns for the lower panel. The green spots indicate the VC positions at the times indicated for the snapshot images in Fig. 2. The velocity $\upsilon$ and the amplitude $|\mathbf{X}|$ of the moving VC are plotted versus time in (b) and (c), respectively. VC orientation reversal occurs at the maximum critical velocity and amplitude values. The extreme minimum value of the local $h_z$ for the upward core orientation and its maximum value for the downward core orientation and the local $m_z$ are shown as a function of time in (d) and (e), respectively. The red and blue colors correspond to the upward ($p = +1$) and downward ($p = -1$) and core orientations, respectively. The vertical lines indicate the times when the extreme values of the local $h_z$ have infinite values, which correspond to the singularity of the gyrofield, Eq. (2), along the lines $m_z = -p$.

**FIG. 1**

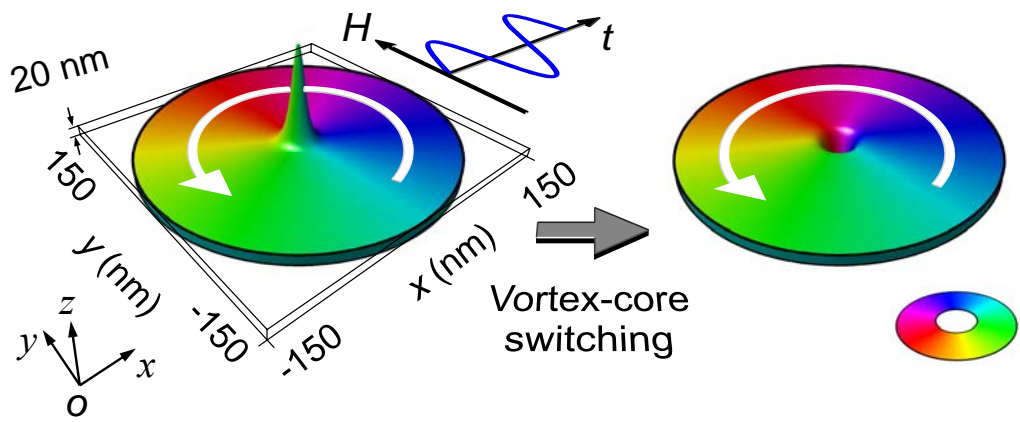

**FIG. 2**

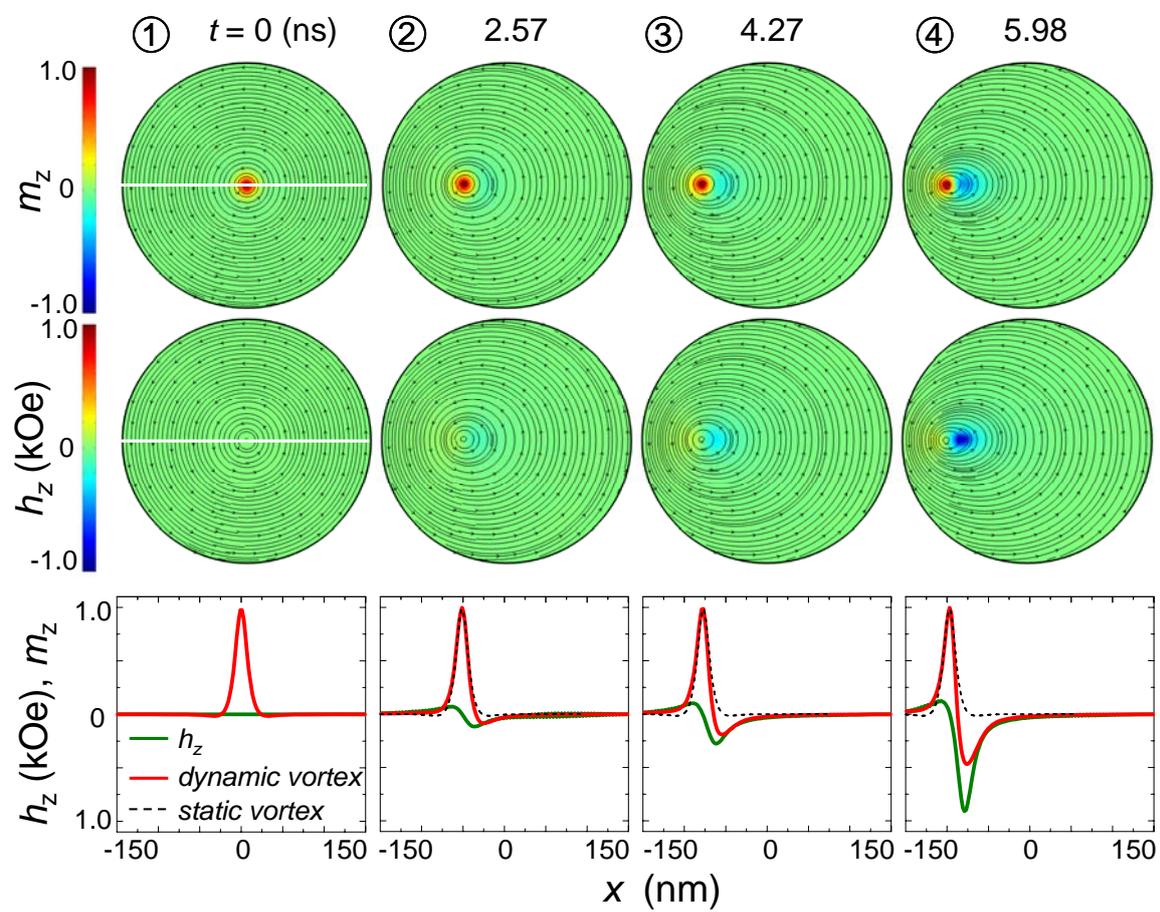

**FIG. 3**

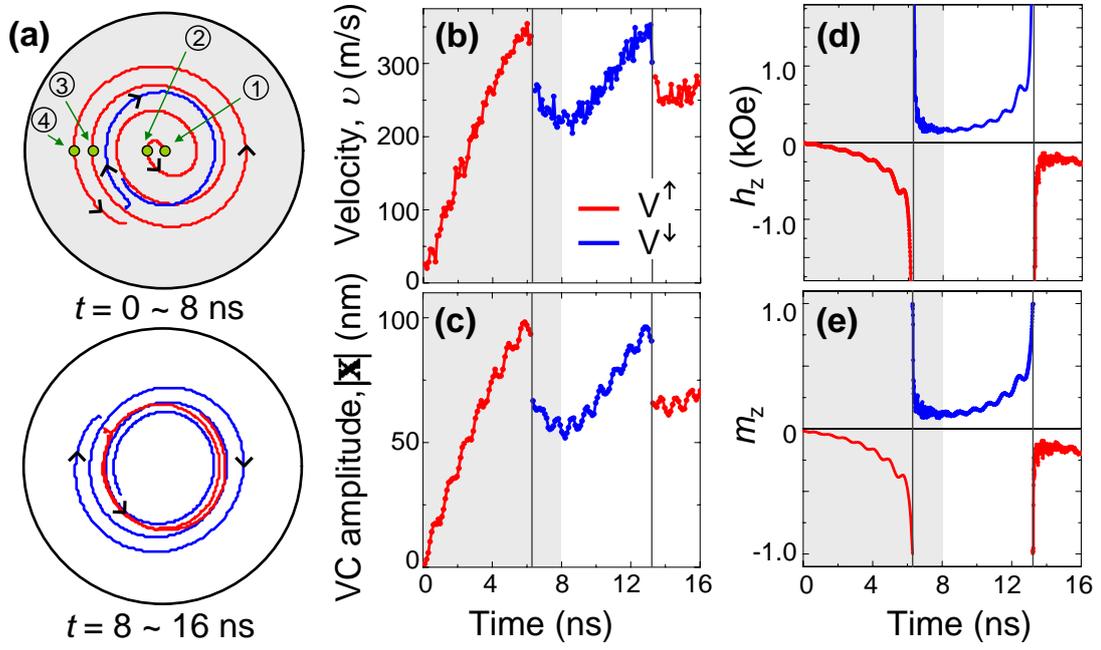